\documentclass[aps,prb,twocolumn,groupedaddress,floatfix]{revtex4}
\usepackage[T1]{fontenc}
\usepackage[french,english]{babel}
\usepackage{amsmath,amssymb,amsfonts,textcomp}
\usepackage{color}
\usepackage{natbib}
\usepackage{calc}
\usepackage{graphicx}

\def\b#1{\mathbf{#1}}

\def\m#1{\mathrm{#1}}

\begin{document}

\title{Wigner islands with electrons over helium}

\author{Emmanuel Rousseau,
\footnote{Present address: Laboratoire
  EM2C{}-CNRS{} - \'Ecole Centrale Paris 92295 Ch\^atenay{}-Malabry (France)}
\email[]{em.rousseau@gmail.com}
Dmitri Ponarin,\footnote{Present address: Chemistry Dept., North Carolina State
                 Univ., Raleigh, NC 27695 (USA)}
Likourgos Hristakos,\footnote{Present address: Science Faculty, American Community 
                 Schools of Athens, 129 Aghias Paraskevis, 152 34 Halandri, 
                 Athens(Greece)} \\
Olivier Avenel, Eric Varoquaux, Yuri Mukharsky}
\affiliation{Service de Physique de l'\'Etat
Condens\'e Centre de Saclay -  91191 Gif-sur-Yvette cedex (France)}

\date{\today}

\begin{abstract}
  We present here the first experimental study of Wigner islands formed by electrons
  floating over helium. Electrons are trapped electrostatically in a
  mesoscopic structure covered with a helium film, behaving as a quantum dot.
  By removing electrons one by one, we are able to find the addition spectrum,
  i.e. the energy required to add (or extract) one electron from the trap with
  occupation number $N$.  Experimental addition spectra are compared with
  Monte Carlo simulations for the actual trap geometry, confirming the ordered
  state of electrons over helium in the island.
\end{abstract}
\pacs{ 67 40}
\maketitle

\section{Introduction} 

Electrons collect on a two-dimensional sheet when they are spread over the
surface of a liquid helium film. This is due to the attractive charge which
appears by polarisation of the helium. Due to the weakness of the image
charge, electrons sit in the vacuum far away from the surface and move freely
at a fixed distance over the film.  As such, they constitute a very clean and
predictable system. 

When temperature is reduced, kinetic energy decreases relative to Coulomb
energy; correlations begin to dominate the electronic structure. Wigner
predicted in 1934 that a phase transition would take place in the infinitely
extended system, leading to the formation of a two-dimensional (2D) electron
lattice.\cite{Wigner:34} Wigner crystallisation into a triangular lattice of
electrons over helium has been observed first by Grimes and
Adams.\cite{Grimes:79,Shikin:89} This phase transition takes place below a
transition temperature that is a function of electronic density. For the
infinite system in the classical limit,\cite{Tanatar:89} in which particles
can be treated individually and which is mostly appropriate for electrons over
helium, the transition occurs when the ratio $\mathit\Gamma$ of the
average Coulomb interaction energy $E_\m C$ to the thermal energy $k_\m B T$
becomes greater than 137. In the quantum case, eventually reached at low
enough temperatures, the Wigner solid is predicted to form when the Brueckner
parameter $r_\m s$, the interelectron distance normalised by the Bohr radius,
$a_\m B$, becomes smaller than 37. Recently, somewhat more complex phases and
ordering transitions have been predicted around that $r_\m s$
threshold.\cite{Waintal:06,Ghosal:07}

An assembly of rectilinear vortices provides another example of Wigner lattice
formation.\cite{Stauffer:68,Campbell:79} Abrikosov's triangular lattice for
vortices induced by magnetic fields in type II superconductors
\cite{Abrikosov:57} was first observed indirectly by neutron diffraction
\cite{Cribier:64} and later visualised directly by electron microscopy imaging
by Bitter decoration of the trapped flux lines with ferromagnetic
microparticles.\cite{Essmann:67}

Also, electrons in semiconductor heterostructures localised by an applied
magnetic field undergo a magnetically induced Wigner transition.
\cite{Lozovik:75} They form a well-studied and well-understood
system, the quantum dot, which has been the object of extensive research for
the past two decades.\cite{Shayegan:99}

\subsection{Confined electron structures}

When confining 2D-electrons to a restricted planar area, the breaking of
translation invariance brings in important changes with respect to the
infinite geometry case, in particular because thermodynamic phase transitions
are suppressed by fluctuations. This problem of 2D ordering in restricted
geometries has been extensively studied theoretically, mostly for parabolic
traps with Coulomb interaction between the electrons, first in the limit where
electrons behave as classical
particles,\cite{Campbell:79,Bedanov:94,Schweigert:95,Koulakov:98}, next in the
computationally more demanding quantum
case.\cite{Egger:99,Filinov:01,Harju:02,Reimann:02}

Finite-size effects become prevalent for $N\lesssim 100$. Such is the case in
the work described below. New features appear that depend on the competition
between the triangular lattice, which takes over for sufficiently large
systems, and the shape and strength of the confining potential, which tends to
suppress it. For hard confining walls and a flat trap bottom, it is predicted
that the ordering is mostly affected close to the boundaries, electrons in the
interior of the 2D island retaining the triangular lattice structure. For
traps with a parabolic confining potential, the particles arrange themselves
in circular shells with widths small compared to the radius of the shell.
Thus, even with very few electrons, ordered structures can appear, the Wigner
molecules.

The direct observation of such structures in restricted geometry has been
achieved only for systems of macroscopic charged
particles,\cite{Saint-Jean:01} and for vortices in
superconductors\cite{Gregorieva:06} and superfluid
helium.\cite{Williams:74,Parts:95}

For electrons over helium, finite islands of electrons can now be realised
\cite{Papageorgiou:05} so that well-controlled experiments can be conducted on
this very clean system, which we report here.  Direct (visual) observation of
the electronic structure cannot be performed in the present experiments: the
geometric arrangement of the electrons in the island must be deduced from
properties such as their escape energy from the island.

\subsection{Wigner molecules}

Contrarily to solid-state quantum dots, for which interactions between
electrons decrease exponentially due to the screening from surrounding
electrodes, electrons over helium are located far from conducting bodies; the
Coulomb interaction is mostly unscreened and gives rise to strong,
long-range, interparticle correlations. Also, there are no nearby impurity, no
effective mass correction; image charges are well-defined. These features
concur to make Wigner islands of electrons over helium an ideal model for the
study of strongly correlated few-body fermionic systems.

The electron states and energies depend markedly on the trap geometry and on
the details of the interparticle interaction, which in turn depends on the
number of electrons in the trap and their arrangement. For traps with steep
walls and flat bottoms, the electrons tend to form a triangular lattice in the
island interior and adjust to the walls at the trap periphery with a
disorganised layer. For circular parabolic traps, electrons order first on
concentric shells following a Hund-type law.\cite{Bedanov:94} These structures
are referred to as Wigner molecules.
 
When fluctuations, either thermal or quantum, are further reduced by lowering
the temperature or by adjusting the electron areal density, the angular positions
of the various shells become locked: orientational order sets in amongst
the previously radially ordered electrons.  Some particularly stable
configurations are found that correspond to ``magic numbers'' of electrons
with high ordering temperatures while other trap occupation numbers lead to
less stable structures and lower ordering temperatures.

\begin{figure}[t]
  \includegraphics[width=80mm]{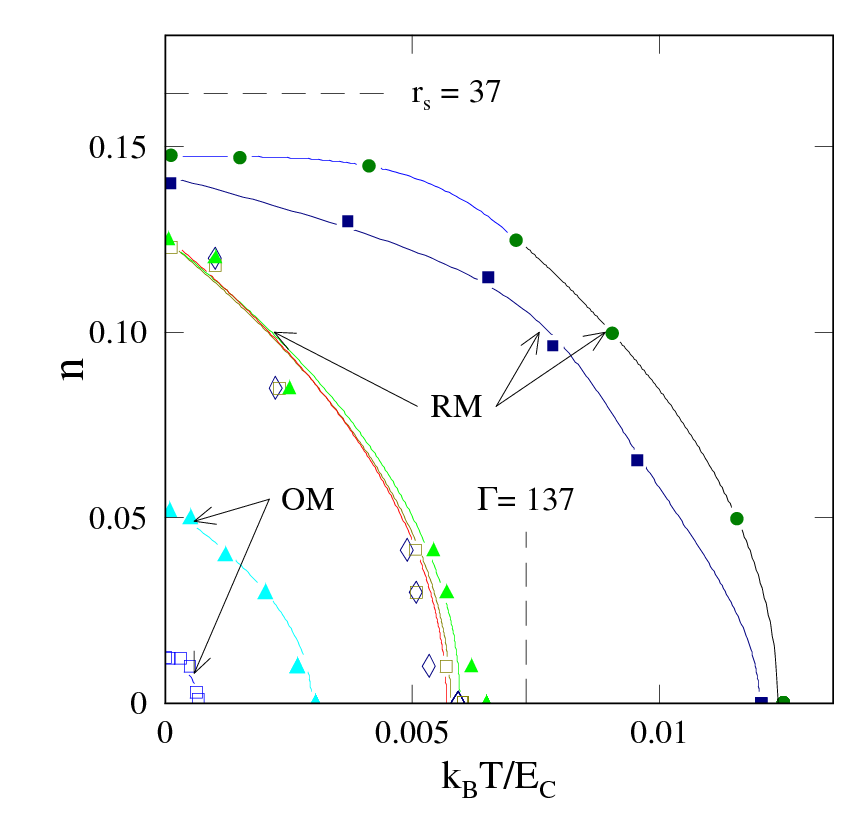}
  \caption{ \label{PhaseDiagram} 
    Radial (RM) and orientational (OM) ``melting'' in Wigner islands sketched in
    the $n$ {vs} $k_\m B T/E_\m C$ plane for small electron occupation numbers
    $N$ - ($\Diamond$) 10 - ($\bullet$) 11 - ($\square$) 12 -
    ($\blacktriangle$) 19 - ({\tiny $\blacksquare$}) 20, following Filinov
    et al.\cite{Filinov:01} The OM boundaries for $N$=10, 11 and 20 are
    very close to the origin and cannot be shown. The horizontal (vertical)
    dash-dash lines indicate the quantum (classical) Wigner crystallisation in
    the infinite system, for $r_\m s = 1/n^2=37$ and $\mathit\Gamma_\infty =
    E_\m C/k_\m B T = 137$ respectively.  }
\end{figure}

\subsection{``Phase'' diagram}

The various ordering processes taking place in 2D-clusters of electrons - the Wigner
islands -  and the formation of radially correlated structures - the Wigner
molecules - have been the subject of a large number of theoretical
studies. Our interest here lies in systems with a small number of electrons,
$N\lesssim 20$, confined in relatively large traps so that the electronic
density is low.   

The behaviour of $N$ electrons freely suspended in vacuum over liquid helium
and confined laterally in a circular parabolic trap by a potential
$\frac{1}{2}m_\m e \omega_0^2r^2$, $m_\m e$ being the bare electron mass and
$\omega_0/2\pi$ the harmonic trap frequency, is described by the following
Hamiltonian
\begin{equation}        \label{Hamiltonian}
  \mathcal H = \sum_{i=1}^N \left\{\frac{\hbar^2\nabla_i^2}{2 m_\m e} 
    + \frac{m_\m e \omega_0^2}{2} r_i^2 \right\} +  
    \sum_{i<j}^N\frac{e^2}{4\pi\epsilon_0|\b r_i -\b r_j|} \; .
\end{equation}  
The terms in curly brackets describe noninteracting electrons in the trap. The
characteristic length $l_0=(\hbar/m_\m e \omega_0)^{1/2}$ associated with
these two terms can be viewed as the spatial extent of the electron
motion. The Coulomb interaction, the last term in Eq.(\ref{Hamiltonian}),
eventually localises and orders the electrons within the trap to distances of
the order of $r_0$ such that $E_\m C = e^2/4\pi\epsilon_0 r_0 =\frac{1}{2}m_\m
e \omega_0^2r_0^2 $.  Following Filinov et al.\cite{Filinov:01}, we can
therefore take the quantity $n=\sqrt 2\, l^2_0/r^2_0$, which represents the
fraction of the trap area actually occupied by the electrons, a dimensionless
measure of the electronic areal density.

The configuration and energies of Wigner islands with $N$ electrons are
described by Eq.(\ref{Hamiltonian}), the eigenstates of which however can be
found analytically only for $N\leq 2$.\cite{Pfannkuche:93} Various
approximation schemes have to be used for higher occupation
number.\cite{Anisimovas:98,Balzer:06,Simonovic:06} Larger electron islands
require the recourse to numerical simulations (see, for a review, Reimann and
Manninen\cite{Reimann:02}).  Interactions with the environment - heat bath and
measuring equipment - are not included in Eq.(\ref{Hamiltonian}). In the
simulations of Filinov et al.\cite{Filinov:01} the temperature comes in as an
input in the Monte-Carlo simulations.

A generalised ``phase'' diagram for the various ordering processes - radial
and orientational - taking place in Wigner islands with few electrons has been
constructed numerically by Filinov et al.\cite{Filinov:01} whose findings are
sketched in Fig.\ref{PhaseDiagram}. These authors characterise the onset of
order by looking at the correlations between electrons both radially and
angular-wise. In the ordered state, they find that $r_0$, introduced above on
dimensional grounds, is quite close to the mean interparticle distance defined
by the first maximum of the pair distribution function. When coming from the
dense, hot (large $n$ or/and $T$) regions of the diagram, where the 2D
electron cluster is in a disordered (liquid or gas-like) state and moving to
more dilute, cooler region, boundary lines are found first to the radially
ordered state (RM), and, moving further in, to complete ordering where the
orientation of the various shells become frozen with respect to one another (OM).

\subsection{Electronic configurations}

As seen in Fig.\ref{PhaseDiagram}, Wigner ordering takes place at low
densities $n$ and low reduced temperatures $k_\m B T/E_\m C$ along boundaries
that differ markedly for various occupation numbers $N$. Some clusters are
more stable than others. Specifically, clusters with $N=$11 and 20 in
Fig.\ref{PhaseDiagram} melt radially at higher $T,\,n$, but orientationally at
much lower values of $T,\,n$.\cite{Filinov:01} The orientationally more
stable clusters are called magic clusters, $N$=10, 12, 16, 19 in
Ref.[\onlinecite{Filinov:01}], and possess a fully frozen structure that is
closest to the Wigner triangular lattice so that the shells cannot rotate on
one another. Those that can, $N$=11, 20, are far less stable orientationally,
but a little more stable radially.

Somewhat surprisingly, these more radially stable configurations turn out to
be even more stable than the homogeneous Wigner crystal itself, which melts in
the classical limit for temperatures above that set by $\mathit\Gamma_\infty
=137$. For the full quantum case, the critical melting density for the
homogeneous system is set by the value of the Brueckner parameter $r_\m
s\equiv r_0/a_\m B =37$.  The Brueckner parameter can be extended to finite
systems by taking the definition of $r_0$ in an electron cluster given above,
which yields $r_\m s=1/n^2$. All boundary lines in Fig.\ref{PhaseDiagram} lie
below the melting curve for the quantum Wigner crystal.

It should be mentioned however that these boundaries do not correspond to sharp
transitions between electronic structures with different types of order.  The
localisation process of the electrons into organised structures takes place
gradually only, as stressed by \citet{Ghosal:07} in particular.  

Also, considerably lower critical values for $r_\m s$, down to 4,
corresponding to much higher critical densities, have been reported for
clusters with few electrons.\cite{Egger:99} However, these findings seem to be
an artifact of the computational scheme and are not confirmed in subsequent
work.\cite{Reimann:00,Reusch:01,Ghosal:07} It thus does appear, as shown in
Fig.\ref{PhaseDiagram}, that the critical density beyond which quantum
fluctuations dominate, destroying the ordered phase of 2D confined structures
remains of the same order of magnitude as for the open-geometry situation when
the number of particles, $N$, is large.

Detailed studies of the ordered configuration of these few-body clusters have
been performed, mostly by numerical simulations, both for systems of vortices
in rotating helium and superconducting
disks,\cite{Stauffer:68,Campbell:79,Schweigert:98} and in quantum dots by
a number of authors, notably Peeters et
  al.\cite{Bedanov:94,Schweigert:95,Szafran:04} in the classical limit, to
whose work we compare our own simulations in Sec.\ref{AdditionSpectra} below.

Visual observations of the ordering of small charged metallic spheres confined
in circular trap on a plane have been performed by \citet{Saint-Jean:01}
Somewhat unexpectedly, their observations of the ground state configuration of
clusters with $N$ up to 30 led to significant differences with the predictions
for some of the structures simulated by more sophisticated Monte-Carlo (MC)
and molecular dynamics (MD)
techniques.\cite{Bolton:93,Bedanov:94,Schweigert:95,Lai:99} They however
agreed with the structures found in older work based on a spatial step-by-step
minimisation of the free energy.\cite{Campbell:79} For instance, the cluster
with $N=15$ is found to order in the ground state with an inner shell of 4
electrons\cite{Campbell:79} against 5 electrons in MC-MD calculations. In
other words, an outer shell with $N=10$ is favoured by the Monte Carlo
approach compared to one with $N=11$ in the spatial minimisation
procedure. Likewise, for a cluster with $N=16$, the observed structure has two
shells, 5 electrons on the inner shell and 11 on the outer. The MC-MD
structure would comprise three shells, one electron at the centre, 5 and 10 on
the two other shells.

The MC approach has been reexamined by \citet{Kong:02} in order to resolve
this discrepancy. These authors paid careful attention to the metastable
states and the saddle paths between them. They confirm the findings of earlier
work with the same computational technique (see also \citet{Kong:03}). Noting
that the energy difference between the true ground state for $N=9$ and 16 - in
their calculation - and the closest metastable state - observed as ground
state by \citet{Saint-Jean:01} is quite small, these authors suggest that
either the experimental situation differs from the one described by the
simulation, or the observed configuration gets consistently stuck in the
metastable state. The precise cause for the discrepancy is not really
uncovered yet although there are hints that electron structures are quite
sensitive to the exact shape of the potential: a departure from a parabolic
confinement potential going as $V \sim r^n$ with $n=2$ to one with $n=2.2$ is
enough to alter the computed structures and restore agreement with
experiment.\cite{Kong:02,Saint-Jean:02} We shall thus consider that MC
calculations carried out with the actual confining potential do lead to
results in agreement with experiments.

\newpage

\begin{widetext} \begin{center}
\begin{figure}[t!]
  \includegraphics[width=120mm]{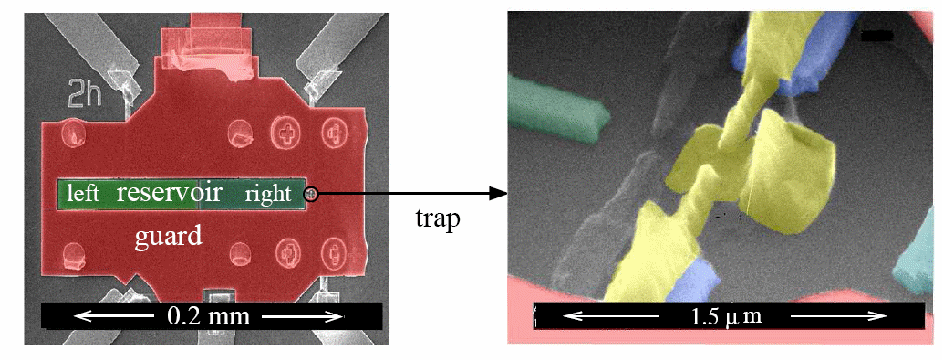}
  \caption{ \label{cell}
    SEM picture of the complete micro-fabricated device (left) and magnified
    view of the electron trap
    (right). 
  }
\end{figure}
\end{center} \end{widetext}
\subsection{Addition spectra}

Addition spectra are built from the energy required to add one extra electron
in the trap. This energy equals the difference in chemical potentials between
clusters with $N$ and $N+1$ electrons:
\begin{equation}        \label{AdditionEnergy}
  \begin{split}
    \Delta^2(N) &= \mu(N+1) - \mu(N) \\
                &= E(N+1) - 2E(N) + E(N-1) \; ,
  \end{split}
\end{equation}
$E(N)$ being the ground state energy of the cluster with $N$ electrons. Once
this last quantity has been determined numerically, the values of the
experimentally more accessible quantity $\Delta^2(N)$ are known.  A study of
the details of this spectrum thus gives a map of the $N$-electron cluster
ground state energies and provides clues about the occurrence of ordering in
the Wigner island.

As it became appreciated that addition spectra provide experimentally
accessible signatures of the onset for the formation of shells in Wigner
molecules,\cite{Tarucha:96} detailed theoretical predictions for various
confinement potentials appeared in the literature (see the reviews by
\citet{Kouwenhoven:01} or \citet{Reimann:02}, and, for more recent
work,  notably \citet{Ghosal:07} and \citet{Guclu:08}).

In the weakly-interacting case, when the last term of the right hand side of
Eq.(\ref{Hamiltonian}) gives a negligible contribution, the quantity
$\Delta^2(N)$ remains small as shells simply fill in and $ \mu(N+1) \simeq
\mu(N)$. When adding one electron gets a new shell started, for $N=3,\; 6,\;
10,\;15,\; \ldots$, the addition energy shows spikes that reveal the shell
structure of the 2D harmonic confinement potential.

As the confinement potential decreases with respect to the Coulomb interaction
contribution, i.e, as the electron density $n$ diminishes, the Wigner
triangular lattice structure eventually forces its imprint on the electron
cluster. A new addition energy signature appears with peaks at $N=3,\; 7,\;
11,\;13,\; \ldots$.\cite{Guclu:08} The crossover occurs around $n=r_{\m
  s}^{-1/2} \sim 0.22$: denser clusters are in a liquid-like state, thinner
ones gradually develop a crystal-like state. As emphasised by
\citet{Ghosal:07} the transition is smooth; addition spectra are expected to
change gradually from shell formations to full spatial ordering. The ``phase''
diagram in Fig.\ref{PhaseDiagram} shows different thresholds for the onset of
order as revealed by the study of pair correlations.\cite{Filinov:01} It
however exhibits the same qualitative trend in the ordering sequence of the
electron islands, which can be studied by measuring the addition spectra.

Our goal in the work reported here is a comparison between the addition
spectra observed in the trap for electrons over helium used experimentally and
the outcome of numerical simulations that we have performed using the known
geometry of the trap. This trap is described in Sec.\ref{ElectronTrap}, the
experimental procedure in Sec.\ref{ExperimentalProcedure}, the obtained
spectra in Sec.\ref{AdditionSpectra}, and the numerical simulations in
Sec.\ref{MonteCarloSimulations}.

\section{The electron trap}
                        \label{ElectronTrap}
   
The trap used in the present work to confine electrons to a 2D island is
simple compared to other traps used for charged particles, which are
usually closed by radiofrequency fields.  Here, the electron motion is
restricted along a plane, on the one side, by the free surface of liquid
helium that presents an energy barrier of $\sim 1$ eV to the electron, on the
other side, by the image charge in the fluid bulk that provides electrostatic
attraction.

At temperatures below 1 K the saturated vapour pressure of helium is very low:
electrons float above the surface of helium in near absolute
vacuum. Unlike solids, liquid helium contains no impurities. Its topological
defects - the vortices - and its elementary excitations - the phonons, rotons
and ripplons - are well known, allowing relatively simple calculations of
interactions between the object in the trap and the
environment.\cite{Fisher:79,Dykman:03,Lea:00} A comprehensive review of the
physics of electrons on liquid helium can be found in
Ref.[\onlinecite{Andrei:97}].

The free electron is attracted toward the helium surface by its image charge
and is repelled by the 1 eV barrier for entering the liquid bulk; it forms
what can be viewed as a one-dimensional hydrogen-like atom. The dielectric
constant of helium is low ($\epsilon=1.057$) and the image charge $Q=e
(\epsilon-1)/(\epsilon+1)$, $e$ being the electron charge, is small. The
characteristic lengthscale analogous to the Bohr radius is $a^*_\m B=4a_{\m
  B}(\epsilon+1)/(\epsilon-1) \approx 76$ \AA, $m_\m e$ being the bare
electron mass, is much larger than the atomic scale. The electron in its
ground state is floating at $1.5 a^*_\m B\approx 114$ \AA\ above the helium
surface. An electric field can be applied externally to press the electron on
the surface and tune both its height and energy. 

If unconfined laterally, the electron moves freely over the surface of bulk
helium. Its mobility is the highest of all condensed matter
systems.\cite{Shirahama:95} The transverse motion of the electron can be
restricted by a system of electrodes located below the helium surface. The
micro-fabricated device used in the present work is shown in Fig.\ref{cell}
and is described in full detail in Ref.[\onlinecite{Rousseau:07}].

The right panel of Fig.\ref{cell} shows the trap where electrons form a Wigner
island.  It is a three dimensional structure micro-machined on a silicon wafer
consisting of {\it i}) a system of electrodes designed to hold the electrons
in a well-defined position over the liquid helium surface, {\it ii}) a single
electron transistor (SET) on a pyramidal island at the centre of the trap used
to detect the presence of electronic charges and, possibly, the excitation
state of the cluster.  The pyramidal shape was made with a 5-angle evaporation
procedure based on the well-known shadow technique.  All electrodes are made
of niobium except the SET which is made of aluminium. More details are given
in Ref.[\onlinecite{Rousseau:07-Thesis}].

The trap and the reservoir, shown in the left panel of Fig.\ref{cell}, are 600
nm lower than the surrounding guard electrode. Both are filled with liquid
helium.  The long rectangular region acts as a reservoir for surface electrons
storage.  If all electrons in the trap happen to become lost, the trap can be
replenished by tapping the reservoir. The guard electrode is made out of a
thick ($\sim $0.25~\textmu m) layer of Nb deposited on an insulating layer of
comparable thickness. This electrode is used to support, by surface tension,
the helium film over the reservoir and the trap ring, so that the liquid depth
is $\sim 0.5$ \textmu m.  The electrode itself and the rest of the sample are
covered by a thin ($\sim $200-400~{\AA}) film of helium, held by Van der Waals
attraction. The bottom of the reservoir is covered by two electrodes, made out
of thin niobium. These separate electrodes control the potentials of the right
and left halves of the reservoir. They are used to shuffle electrons from one
side to the other. The resulting change in capacitance provides a mean of
gauging the total mobile charge contained in the reservoir.

\begin{figure}[tb]
  \includegraphics[width=80mm]{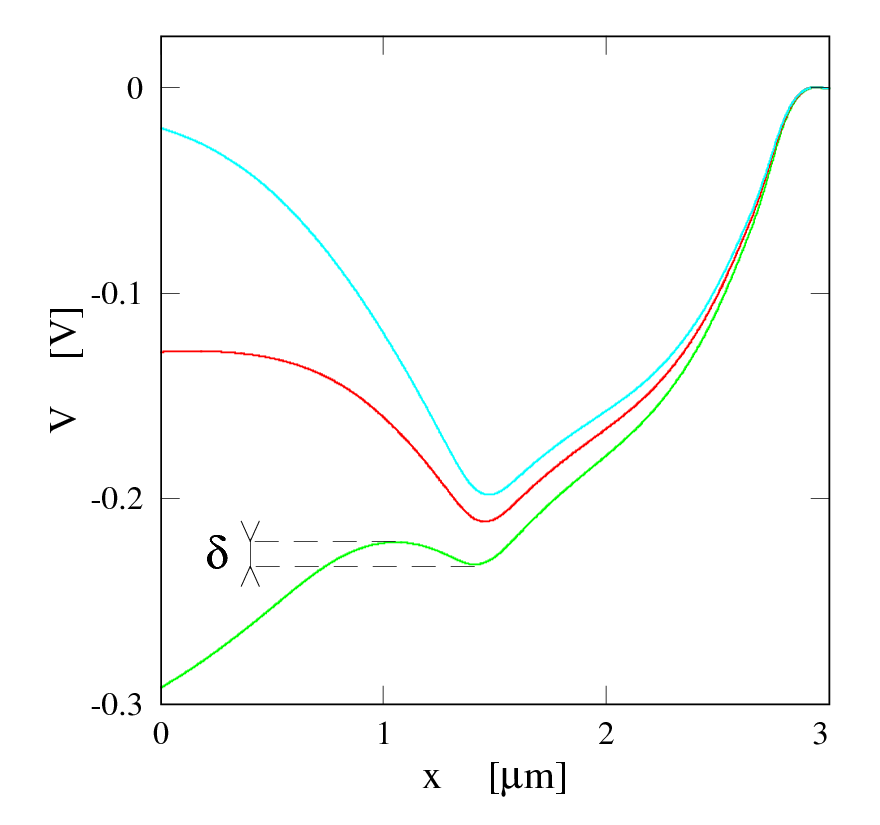}
  \caption{ \label{Potential} Profile of the electrostatic potential in the
    plane of symmetry of the trap for $V_\m{SET} =0.5,\;0.2 \; {\mbox{and }}
    0$ V from bottom to top. Distances along the $x$-axis are in microns from
    the edge of the reservoir. The top of the barrier is in the gorge. The
    difference between the top of the barrier and the minimum define the
    confining potential $\delta$.  }
\end{figure}

A gorge through the guard electrodes connects the reservoir to the
trap. Usually, the guard electrode is biased to negative potential relative to
the SET. A potential barrier thus forms at the gorge, separating the trap and
the reservoir. As the right reservoir electrode protrudes inside the gorge,
the potential on this electrode strongly affects the height of this
barrier. Typical potential profiles formed by these electrodes are shown in
Fig.\ref{Potential}. All electrodes are capacitively coupled to the SET island
and can be used as gates.

The geometric radius of the trap is 3 {\textmu}m, but its effective radius is
significantly smaller and depends on the configuration of the electrostatic
potentials on the various neighbouring electrodes.  The actual potential well
is neither parabolic nor axisymmetric. Both shape and depth change with the
voltage applied to the electrodes. It becomes shallower when the lability
point, at which the confinement completely disappears, is approached. To set
numbers, a trap frequency of $\omega_0/2\pi$ of $\sim$ 40 Ghz represents a low
estimate when the trap holds few electrons only. The extension of the
ground state wavefunction of non-interacting electrons is then
$l_0=(\hbar/m_\m e \omega_0)^{1/2}\sim 21.5$ nm. The mean distance between
interacting electrons $r_0$ fixed by the Coulomb energy and the parabolic
confining energy is 200 nm, the average density parameter $n$, 0.016, the
Coulomb energy $E_\m C$, 83.5 K. At a temperature of 0.2 K,
$\mathit\Gamma^{-1} = k_\m B T/E_\m C = 0.0024$. Referring to the ``phase diagram'' in
Fig.\ref{PhaseDiagram}, the electron island lies well into the radially
ordered phase region, and, moreover, in the classical part of that region.

Such electrostatic traps have already been constructed and operated by    
\citet{Glasson:05} Building on the experience gained in their work, we have
improved the design in two ways: 1) the size is made
smaller; the electron is better confined; 2) the central electrode
protrudes from the bottom with a pyramidal shape in order to increase the
confinement and the coupling with the electrons (in Fig.\ref{cell}, right
panel), thus enhancing the detection sensitivity.

\begin{figure}[tb]
  \includegraphics[width=80mm]{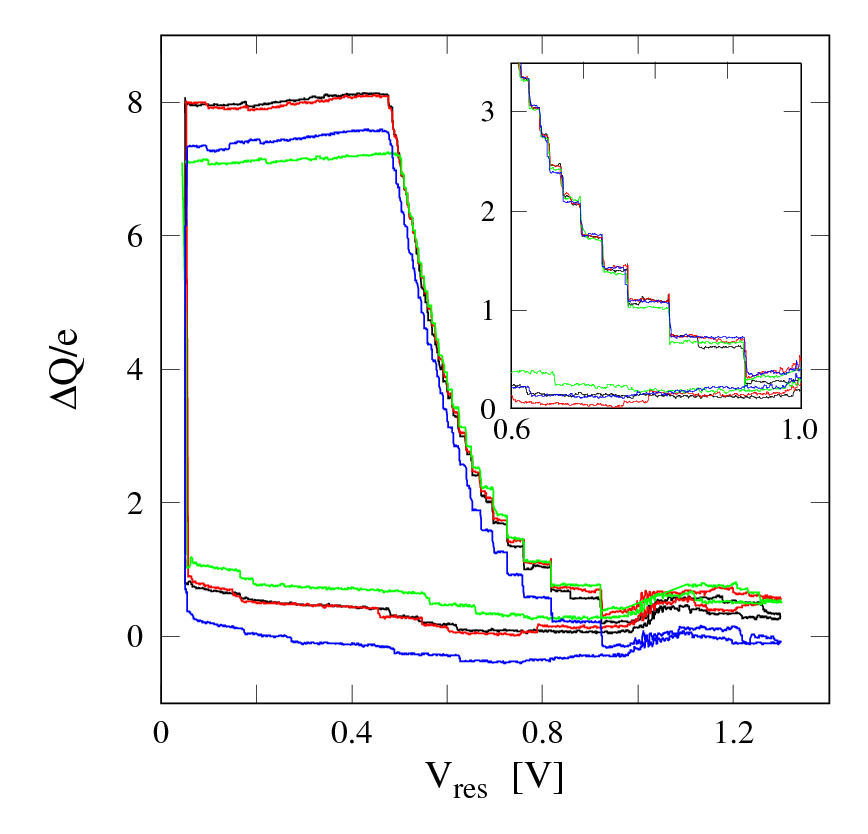}
  \caption{ \label{SETcharge} Reduced charge on the SET island in terms of the
    potential of the reservoir electrode for several sweeps of $V_\m
    {res}$. The SET potential is +0.6 V, the guard electrode potential, 0. The
    inset shows the final steps of the staircase for low occupation numbers
    down to zero, corrected for baseline drift.}
\end{figure}

\begin{figure}[tb]
  \includegraphics[width=74mm]{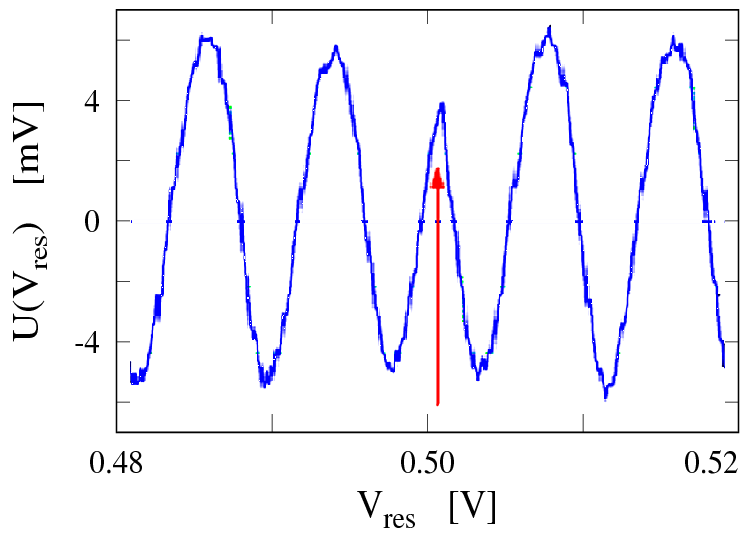}
  \caption{ \label{SETsignal} Variation of the SET output signal $U$ when the
    reservoir voltage, $V_{\m{res}}$, is swept. The Coulomb blockade
    oscillation is interrupted by a change of the charge in the SET island at
    $\sim 0.5$ volt as marked by the arrow.}
\end{figure}

\section{Experimental procedure}
                        \label{ExperimentalProcedure}  

\subsection{Electron seeding and monitoring.}

We seed the electrons on the helium surface by igniting a corona discharge in
a small chamber separated from the rest of the cell by a metallic grid with
mesh of a few tens of microns.  Before the
discharge, the cell is heated to a temperature $\sim $1.1~K at which the
vapour pressure becomes high enough.  A high vapour pressure is
required both to ignite the discharge and to thermalise the electrons so that
their energy is lower than the energy barrier of 1~eV needed to penetrate into the
liquid.  The discharge is ignited by applying $\sim $\,-500 volts to a wire
terminated in the middle of the discharge chamber.  The typical discharge
current is $\sim $0.1-0.2~\textmu A.  The presence of electrons is easily
detected by applying $U_\m{exc}\sim $100 \textmu volt of 100~kHz voltage to the
right reservoir electrode and measuring the voltage induced on the left
electrode with a lock-in amplifier.  When electrons appear on the surface, the
signal changes by 1-20~10$^{-8}$ volt rms.

After the electrons have been generated and have scattered over the cell a
negative potential is applied to the guard electrode while a positive
potential is applied to the reservoir (the potential difference is typically
between 0 and +1 V during the discharge). These confining potentials localise
the electrons primarily over the reservoir and the ring shaped trap.

At low temperature, the electrons over the thin Van der Waals film become
localised while electrons over the reservoir and the trap ring, floating much
further away from the solid substrate, remain mobile.  The thickness of the
helium film covering the reservoir and the trap depends somewhat on the amount
of helium in the cell. As the level of helium decreases, the film thickness
also decreases. When the helium level drops to $\sim 5$ mm below the sample surface,
only a Van der Waals film remains at the centre of the reservoir. In a cell
with a small volume, it is rather difficult to precisely meter the liquid level
as significant amounts of helium can remain trapped in the fill line by surface
tension and fountain pressure. To alleviate this problem we have increased the
volume of the cell below the sample.  We determine the amount required to fill
the cell up to the chip level by measuring the capacitance ($~\sim$ 1 pF)
between two pads on the chip as a function of the volume of helium condensed into
the cell.  When helium liquid starts covering the chip, the capacitance increases. We
then empty the cell and refill it with a quantity of gas corresponding to that
just before the capacitance increase.

\subsection{SET readout}

The single-electron transistor (SET) located at the bottom of the trap
operates as a sensitive electrometer and quantum
amplifier\cite{Devoret:00,Likharev:03} to detect the presence of electrons in
the trap and, possibly, their quantum state.  This SET is current-biased and
polarised close to the Josephson quasiparticle peak. 

We record the Coulomb blockade oscillations obtained by sweeping the reservoir
potential.  The voltage across the SET, $V_\m{S}$, is modulated according to
the total charge on its island, which reads
\begin{equation}
  q_\m I = \sum_i C_i U_i + \sum_j q_{0,j}  \; .
\end{equation}
The first sum is over all the conductors in the system with capacitance to
the island $C_{i}$ and potential $U_{i}$. The second sum runs on the charges
induced on the island by free charges in the system, such as charges or
dipoles in the substrate and other electrons over helium. 

Voltage $V_\m{S}(q_\m I)$ across the SET is a periodic function of charge
$q_\m I$ with period $e$, the electron charge. However, it also depends on the
bias current and the electronic configuration in the cell, and varies from run
to run.  We determine the functional dependence of $V_\m S(q_\m I)$ on $q_\m
I$ experimentally by sweeping the potential $U_\m e$ of one of the electrodes,
usually the one that is swept in subsequent measurements. We then select a
portion of the sweep during which the background charge distribution did not
change so that several cycles of the function $V(q_\m{I})$~=~$V(U_\m{e})$ can
be superimposed by transforming $U_\m{e}$ modulo $P$, where $P$ is an
appropriately chosen period. After the modulo transformation the selected
piece of data is averaged and interpolated by a smoothing spline function
$\mbox{spl}\,(U_\m{e})$. The rest of the data is fit piecewise with the function
$A\, \mbox{spl}\,(U_\m{e}+q P)$, where the amplitude $A$ and the phase $q$ are
fitting parameters.  The amplitude has to be taken as a free fitting parameter
because the experimentally observed $V(U_\m{e})$ varies somewhat in amplitude
with $U_\m e$.

The phase $q$ is the charge, expressed as a fraction of $e$, induced on the
SET island by free charges in the system.  The expected value is easy to find
from the reciprocity principle: it is equal to that induced at the location of an
electron when the SET island is biased with unit potential and all the other
conductors in the system are grounded. We have calculated this potential
numerically, using the actual 3D geometry of the trap inferred from SEM
photographs, to obtain a value for the charge $q$ of 0.5~$e$. 

To reduce the noise due to fluctuations of the {\it{dc}} voltage across the
SET we apply a low-frequency (80-150~Hz) modulation with amplitude $\sim$ 100
\textmu volts to the guard electrode. The amplitude of the detected signal
is proportional to the derivative of the periodic function $V_\m{S}(q_\m{I})$.

\begin{figure}[tb]
  \includegraphics[width=80mm]{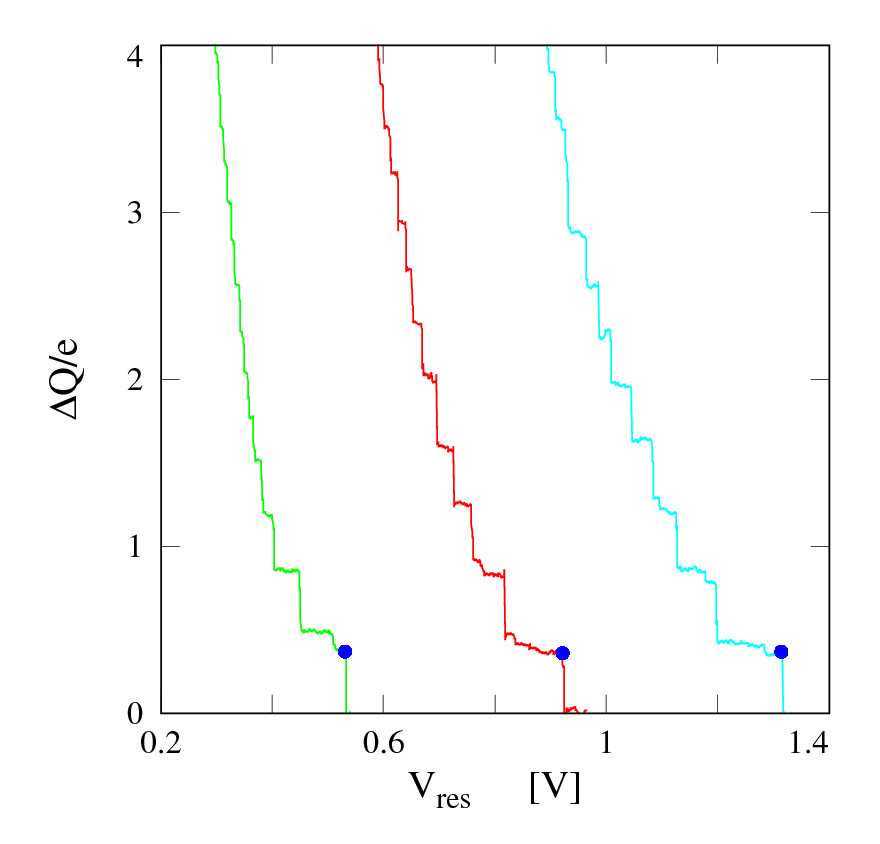}
  \caption{\label{SETphase} SET phase curve variation for the values of the
    potential on the SET $V_\m{SET}$ of 0.3, 0.5, and 0.8 volt from left to
    right. The dots mark the escape the last electron from the trap.  }
\end{figure}

\begin{figure}[tb]
 \includegraphics[width=80mm]{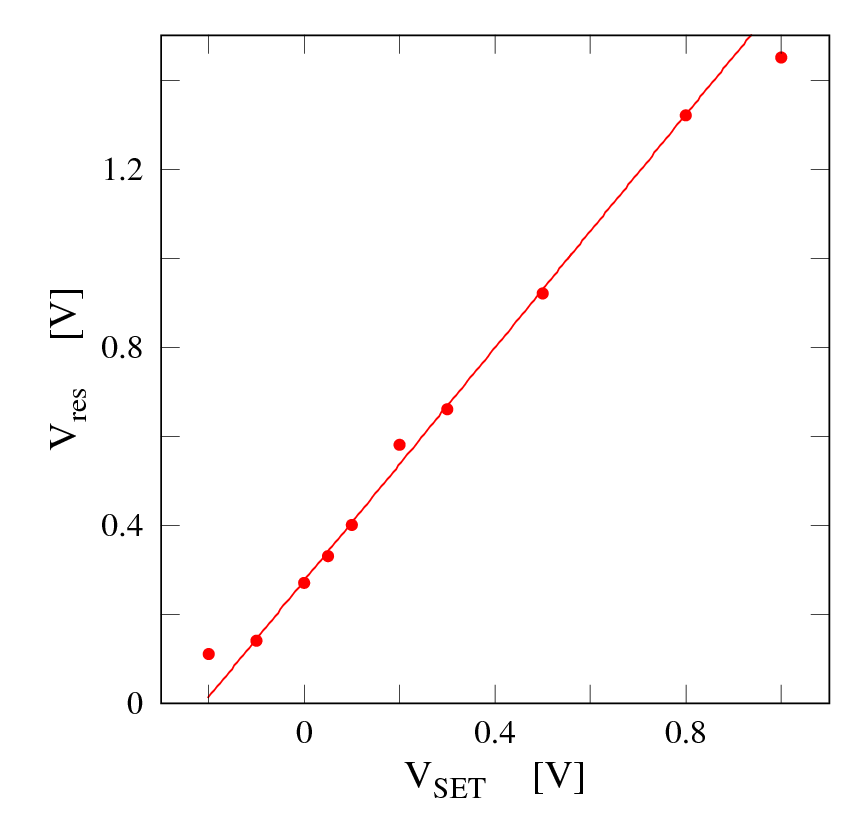}
  \caption{ \label{ContactPotential} 
    Potential on the reservoir for which the last electron leaves
    the trap. The intercept with the $x$-axis yields the effective contact
    potential between aluminium and niobium, found to be $0.206 \pm 0.005$ V.
  }
\end{figure}

\subsection{Trapping electrons}

After electrons are seeded, we let the system cool down from 1~K and proceed
to load the trap with a given number of them. Typically, at this stage, the
guard electrode is biased to a negative potential, between $-0.1$ and
$-0.5$~volt, and the SET is biased to a positive potential between 0 and $0.5$
volt. First, we fill the trap with electrons by lowering the voltage applied
to the reservoir electrodes. The electrons are repelled towards the trap. Care
should be taken not to lower the voltage too much, since this can cause
irreversible loss of electrons. The values of the electrode potentials are
determined by trial and error. Usually, at least one of the reservoir
electrodes stays more positive than the guard, although we found that we can
keep the electrons even with both electrodes more negative than the
guard. Often, the charging potentials have to be lowered in the course of the
experiment as the electron reservoir gradually gets depleted.

We have tried loading the trap in a controlled manner, i.e. with a given number of
electrons, but found that the electrons usually enter in large, unpredictable
bunches, even when the potentials of the reservoir electrodes are swept
slowly. This is contrary to the results reported by \citet{Glasson:05}
The reason for this difference is unclear.

After the trap is charged, we start sweeping the potential of the right
reservoir electrode $V_\m{res}$. When the right reservoir voltage is low, the
barrier is high and the ring remains full of electrons.  Sweeping up this
voltage reduces the barrier height. When the barrier becomes low enough, electrons
start leaving the ring. This escape suddenly changes the charge on the SET
island. The SET island charge variation manifests itself as a phase jump in
the Coulomb blockade oscillations (see Fig.\ref{SETsignal}, in which the arrow
points to the phase jump).

Referring to the traces in Fig.\ref{SETcharge}, the SET detects a sudden
change in the charge at $V_\m{res}\sim $0.48~volt.  Since the SET measures
charge modulo $e$, we cannot determine directly the absolute value of the
total charge but we know that the trap flooded with electrons has started to
empty.

In this initial phase, electrons leave the trap in such a way that individual
escapes cannot be resolved. But, as $V_\m {res}$ is raised further and the
barrier height decreases, clear step-like jumps become visible. Each electron
leaving the trap changes the induced charge on the SET island.  The corresponding
jump of the SET island charge, $\Delta Q$, is plotted in Fig.\ref{SETcharge}
against $V_\m {res}$. Its amplitude depends on the number of the electrons
left in the trap and amounts to $\sim 0.4 e$ when few electrons only are left. This
value compares well with that obtained from finite-element calculations of the
electrostatic field using the known trap geometry, $\sim 0.5 e$: electrons
escape from the trap one by one.

Finally, at 0.93~volt the last electron leaves the
trap. Between 0.82 and 0.93~volt there is only one electron left
in the trap. The next step to the left corresponds to two electrons in the
trap, and so on.

The variation of $V_\m {res}$ between the jumps depends on the Coulomb
repulsion between the electrons. Stairs length and height increase when the
number of electrons in the trap decrease. Length increase means a larger
reduction of the barrier to extract one electron (electron-electron
interactions decrease). Height increase means that the leaving electron
induces a larger charge on the island, i.e., that the leaving electron is
closer to the centre of the island.

These results are similar to the results obtained at the Royal Holloway in
London.\cite{Glasson:05} The main difference lies in significantly better
defined steps, which are both ``higher'' and ``longer'' in the present
work. This is due to better coupling with the pyramidal SET, compared to
flat-island SET used in London, and to the smaller size of the trap, leading
to stronger repulsion between the electrons. The positions of the steps are
more stable in our experiments as well, as seen in the inset of
Fig.\ref{SETcharge}, and amenable to precise quantitative analysis.
 
\section{Experimental addition spectra}

                        \label{AdditionSpectra}

\begin{figure}[t]
  \includegraphics[width=80mm]{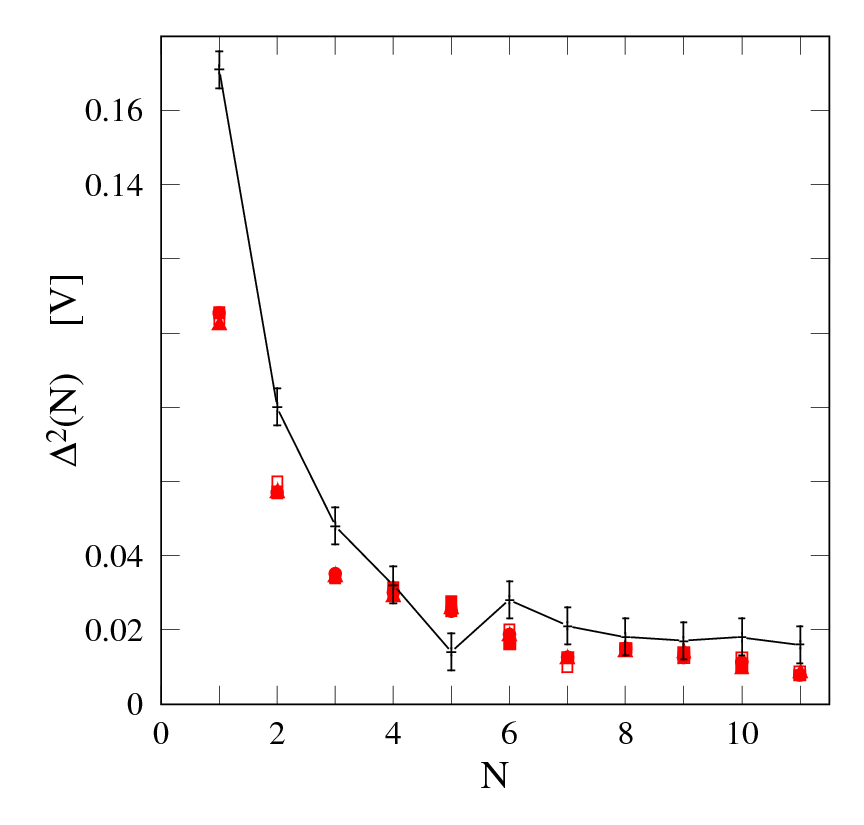}
  \caption{ \label{AdditionSpec2} 
    Addition spectrum for $V_\m{SET}$=0.5 V
    ($V_\m{SET}$=0.706 V in the simulation).  Black crosses are simulation
    results. Error bars represent the uncertainty on the fit
    parameters. Empty and filled squares, and circles are experimental 
    data for different runs and fall nearly on top of one another.
  }
\end{figure}

\begin{figure}[t]
  \includegraphics[width=80mm]{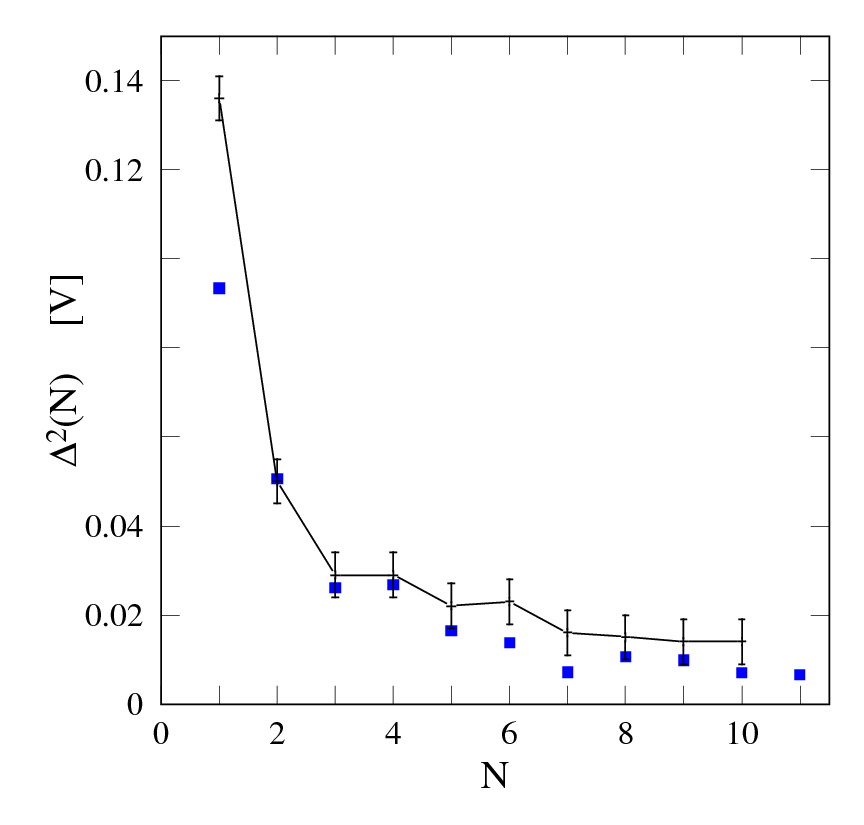}
  \caption{ \label{AdditionSpec1} Addition spectrum for $V_\m{SET}$=0.3 \
    (experimentally) ($V_\m{SET}=0.506$ V in the simulations). Crosses are
    simulation results. Error bars represent the uncertainty on the fit
    parameters. Squares are experimental results }
\end{figure}

In the following, we shift our attention from SET{}-phase variation signals to
addition spectra, which are inherently more reproducible for the following
reason. For a given set of external parameter values, such as the potentials
applied to different electrodes, the potential for which one electron leaves
the trap can be shifted due to modifications in the distribution of trapped
charges in the substrate.  This spurious effect is more likely to occur
shortly after a new corona discharge, which requires a ``high'' temperature
(1{}-1.2 K) and involves a high voltage. As addition spectra are the
difference between $V_\m{res}(N-1)$ and $V_\m{res}(N)$, namely $\Delta\mu(N) =
V_\m{res}(N) - V_\m{res}(N-1)$, potential drifts are removed.

We note that $V_\m{res}$ is modified by the contact potential
between niobium and aluminium in the cell. This potential is determined in the
experimental setup as follows. We read in Fig.\ref{SETphase} the value of the
right reservoir potential for which the last electron leaves the trap for
different potentials applied to the SET (i.e. different sizes of the
trap). These values are then plotted against $V_\m{SET}$ in
Fig.\ref{ContactPotential}. The experimental points fall on a straight line
that cuts the $x$-axis for $V_\m{res}-R = -0.206 \pm$ 0.005 V. This value
represents, to a weak correction due to the image charge of the remaining
electron, the contact potential between niobium and aluminium. It must be
taken into account to obtain the true value of the potential that acts on the
electrons.
 
We now turn to the staircase-like addition spectra themselves. We plot the
stair length in terms of the number of electrons, i.e., the right reservoir
potential variation necessary to extract one electron. This gives the energy
profile in terms of electron number.  Experimentally observed addition spectra
are given in Figs.~\ref{AdditionSpec2} and \ref{AdditionSpec1} for two trap
sizes corresponding to $V_\m{SET}$=0.5 and 0.3 volt respectively.

\begin{figure}[tb]
  \includegraphics[width=80mm]{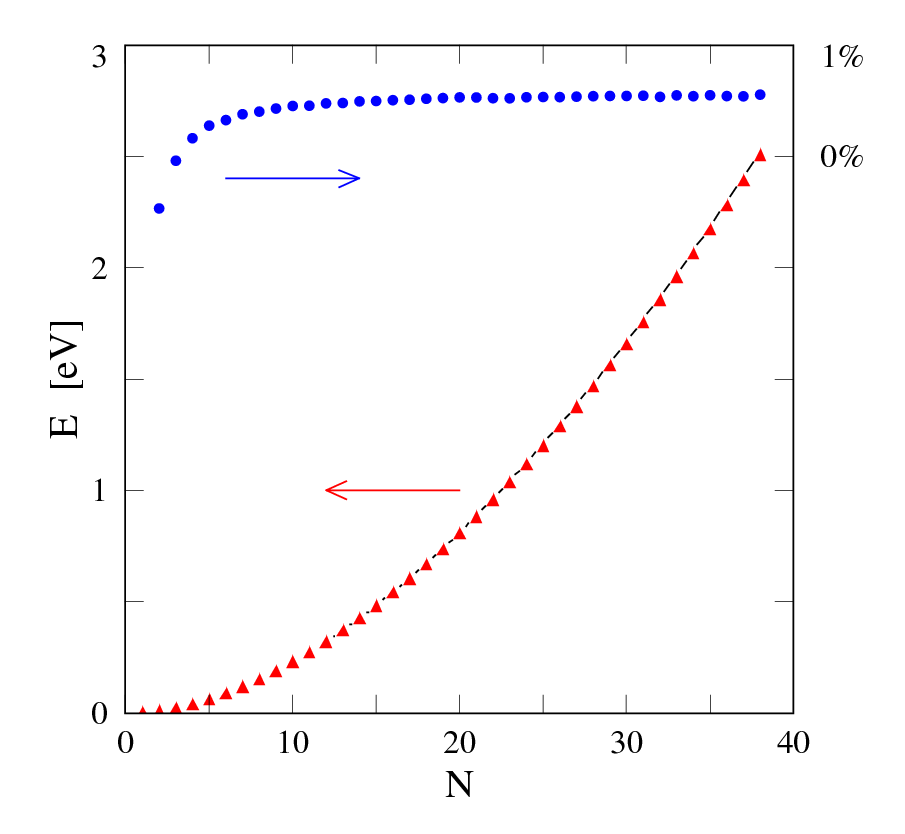}
  \caption{ \label{EnergyChange} Energy change with number of electrons in a
    parabolic trap: ($\blacktriangle$), our results based on a discrete
    parabolic profile, plain line, Kong et al.'s results \cite{Kong:02};
    ($\bullet$, vertical scale on the right),
    relative difference between the two calculations. }
\end{figure}

\begin{figure}[tb]
  \includegraphics[width=85mm]{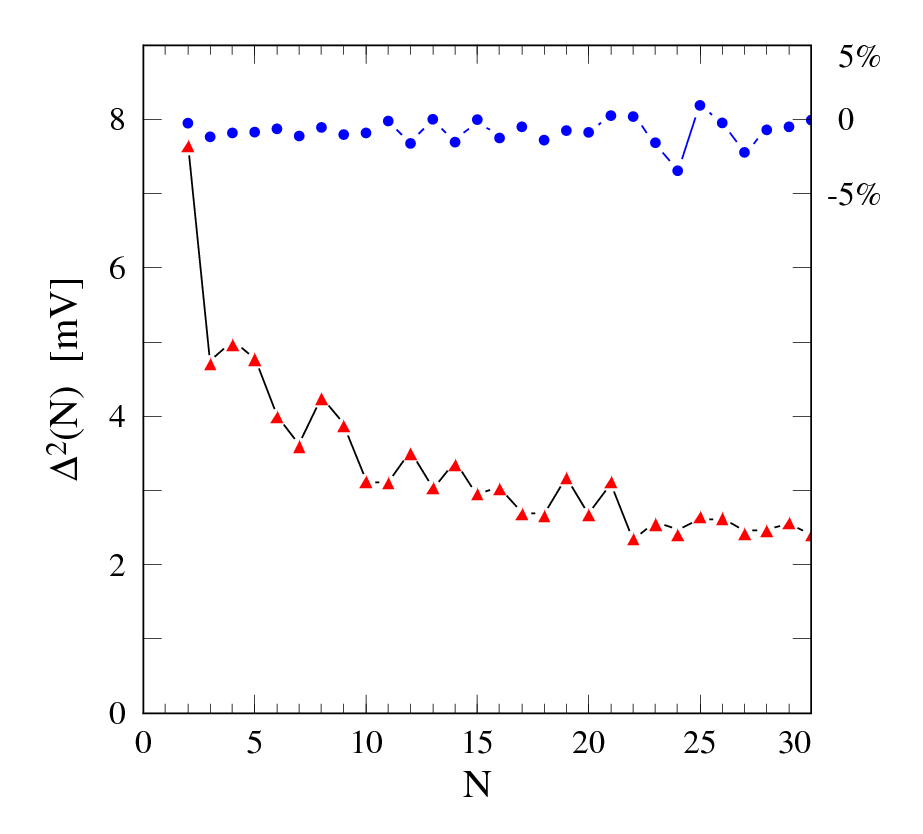}
  \caption{ \label{AdditionSpecTheor}
    Computed addition spectra in this work and that of 
    \citet{Bedanov:94}, confirmed by \citet{Kong:02},($\blacktriangle$), at the
    bottom. The relative difference in \% between these calculations are plotted at
    the top, ($\bullet$, vertical scale on the right).  
  }
\end{figure}

\section{Monte-Carlo simulations}
                        \label{MonteCarloSimulations}

In order to interpret the details of these experimental results, we have
carried out Monte Carlo simulations of the addition spectra that correspond to
the precise shape of the confining potential well in the experimental trap.

These MC simulations are based on the procedure described by Bedanov and
Peeters.\cite{Bedanov:94} In the low density limit, which is the regime
attained here, the electrons occupy a small fraction only of the total area of
the trap. They are (mostly) distinguishable and can be treated as classical
particles. This simplification is also made by Peeters et al.

Our trap is not parabolic but its profile can be determined by finite elements
calculation based on the known geometry (see Fig.\ref{Potential}). Indoing so,
we have attempted to reproduce as accurately as possible the pyramidal shape
of the SET island. We used in the MC simulations values of the confining
potential profile computed for a finite number of nodes whereas Peeters et al.
use an analytic form. To find the electrostatic energy of an electron at
coordinate $\b r_i$ in the trap, we fit the potential profile around $\b r_i$
by a 2D parabola between the three closest nodes of the calculated profile.

In order to check our computational procedure, we have reproduced Bedanov and
Peeters'\,results for a parabolic trap. We discretise the parabolic potential
with the same number of nodes as with the actual potential for our cell. We
then perform the same MC simulations to find the energy and the configuration
of a known number of electrons in the trap. Our results, shown in
Fig.\ref{EnergyChange}, fall very well in line with the published results,
those of Ref.[\onlinecite{Bedanov:94}] and the more recent results of
\citet{Kong:02} The small systematic difference of $\sim$~0.75 \% seen in
Fig.\ref{EnergyChange} may be due to the discretisation of the confinement
potential. We believe that it does not affect the global results and that our
simulation does find the true ground state configuration.

To make sure that our simulations are not noisier than Peeters et al.'s, we
compare addition spectra. Once again, our results are in very good agreement
with those in the literature, as shown in Fig.\ref{AdditionSpecTheor}.

\begin{figure}[tb]
  \includegraphics[width=70mm]{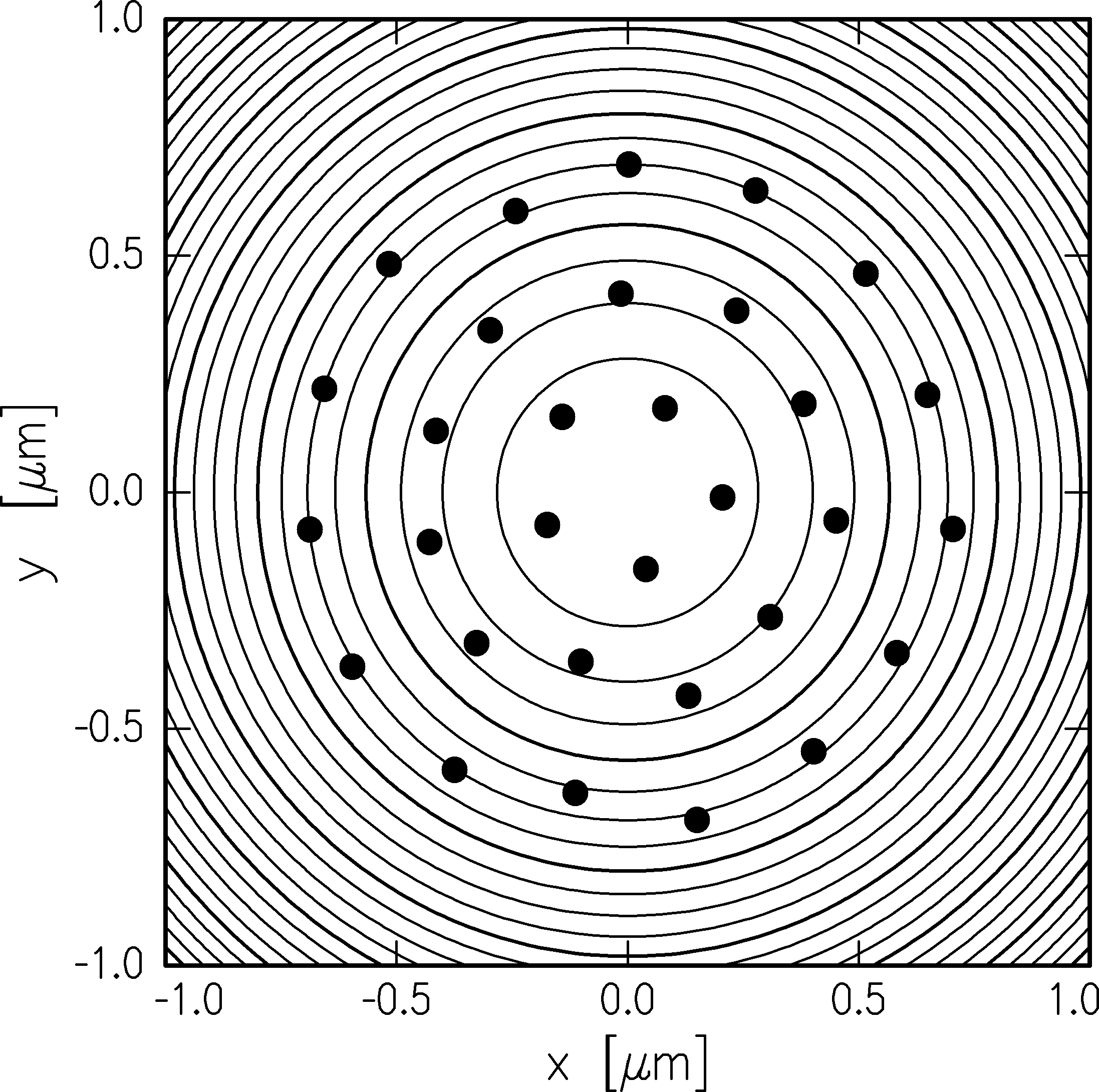}
  \caption{ \label{Molecule}
    Wigner molecule in a parabolic trap with 31 electrons. Lines
    represent iso{}-potential curves, the minimum of the parabola is centred on
    (0,0). Stars represent the electrons. The ground state configuration is the
    configuration (5,11,15) as in \citet{Kong:02}
  }
\end{figure} 

Finally, we compare electronic configurations. The ground state of the
configuration with 31 electrons is one of the most difficult to find due to
the closeness of the first metastable state (the difference is only
0.004\%). The result, shown in Fig.\ref{Molecule}, is once again in agreement with
Peeters et al.'s. We thus are quite confident that our MC
simulations lead to correct configurations and energies.

\begin{figure}[t]
  \includegraphics[width=70mm]{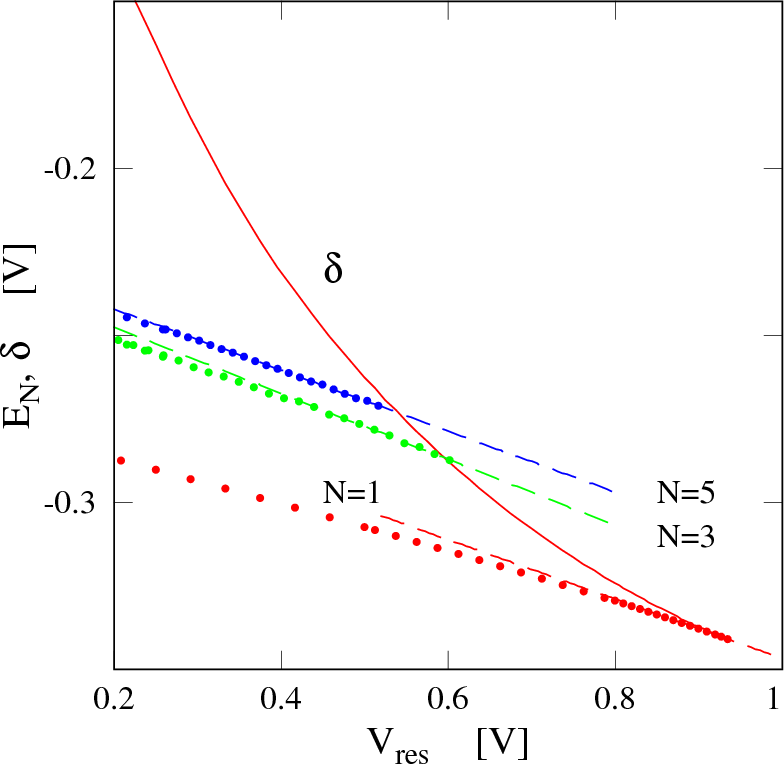}
  \caption{ \label{ElectronEnergy} Evolution of the confining potential
    ${\delta}$ in terms of the potential applied on the reservoir electrode
    $V_\m{res}$. Evolution of the average energy per electron for 1 electron
    and 5 electrons in the dot.  Crosses are Monte{}-Carlo simulation
    results. The plain line is a linear extrapolation of the Monte-Carlo
    results as explained in the text. The crossing point between ${\delta}$
    and this average energy gives the potential for which an electron leaves
    the trap.  }
\end{figure}
 
We now turn to the case of addition spectra in the actual (non-parabolic)
trapping potential, for which we need the electron energy. This energy has two
parts: an electrostatic part and the repulsion due to the Coulomb interaction.
The electrostatic part can be attractive if electrons are in the trap or
repulsive if the electrons are on the other side of the barrier (see
Fig.\ref{Potential}). The electron energy reads, discarding the kinetic energy
term in Eq.(\ref{Hamiltonian}):
\begin{equation}        \label{ClassicalElectronEnergy}
  E(N,V_{\m{res}}) = -\sum_{i=1}^N eV(\b r_i) + 
    \frac{e^2}{4\pi\epsilon_0} \sum_{j<i}^N \frac{1}{|\b r_i-\b r_j|}  \; ,
\end{equation}
where $V(\b r_i)$ is the electrostatic energy, obtained by a finite elements
calculation, $\b r_i$ the $i${}-th electron coordinate, and $e$ the electron
charge.  The second term represents the Coulomb energy between electrons. The
bulk part of the energy comes from the trapping; only around 10 \% of the total
energy in Eq.(\ref {ClassicalElectronEnergy}) is due to the Coulomb interaction.

An electron leaves the trap when its energy overcomes the confining potential
$\delta$, which is the energy difference between the minimum and the energy at
the barrier (see Fig.\ref{Potential}). That is, a given configuration is
stable as long as the average energy per electron remains lower than the
confining potential:
\begin{equation}
  E_\m N = \frac{\textstyle E(N,V_{\m{res}})}{\textstyle N} \leqslant \delta \; . 
\end{equation}
This assumes that electrons have identical energies.

We first compute the evolution of the confining potential $\delta$ in terms of
the potential applied to the reservoir. The contact potential must be taken
into account: when the potentials applied to the SET electrode and to the
reservoir electrode are $V_\m{SET}$ and $V_\m{res}$, the potentials used in
the simulations are $V_\m{SET}+0.206$ V on the SET electrode and $V_\m{res}$
left unchanged on the reservoir electrode. For $N$ electrons in the trap, we
compute the total energy for different values of the reservoir potential. When
the barrier too low - i.e. when the reservoir potential is too close to the
threshold when the electron is about to escape - the Monte-Carlo simulations
fail to find the ground state. The starting temperature in the simulations is
too high and allows electrons to escape readily over the barrier. Using a
lower starting temperature to circumvent the problem does not lead to the real
ground state of the configuration. To resolve this issue we calculated the
energy for values of the reservoir potential slightly lower than the escape
threshold, for which the simulations did find the ground state, and
extrapolated the results linearly to higher values of this potential.

Figure \ref{ElectronEnergy} shows the evolution of the confining potential and the
average energy per electron for N=1, 3, and 5 in terms of the reservoir
voltage. An electron leaves the trap when these curves intersect.

Figures \ref{AdditionSpec2} and \ref{AdditionSpec1} show the comparison
between the outcome of these calculations and the experimental results for
respectively $V_\m{SET}$=0.5 and $V_\m{SET}$=0.3 V. The size and the shape of
the trap depend on $V_\m{SET}$ and the results are significantly different.
Different potential sweeps following a given corona discharge are shown on the
same graph in Fig.\ref{AdditionSpec2}: the experiment gives quite reproducible
results. The scatter on the experimental points is less than the uncertainty
on the simulations.  The observed addition spectra are quite reproducible as
long as the cell is kept cold, below 1 K. When it becomes necessary to
replenish the electron reservoir, the temperature is raised and a new corona
discharge ignited. Then the distribution of stray charges changes and the
addition spectra fine structure, which is quite sensitive to the potential
profile, also changes.

The addition energy is well predicted for $N > 3$. For very few
electrons ($N\leqslant 3$), the Monte-Carlo simulation yields a larger
addition energy.

A recurrent feature of the observed spectra is a peak for $N=6$ followed by a
trough for $N=7$. Referring to the work of \citet{Guclu:08}, this indicates that
electrons order in shells and not on a triangular lattice,
which would give a peak at $N=7$. The corresponding trap frequency is $\sim
60$ GHz and the electronic density $n\simeq 0.017$. These observations
correspond to the phase diagram shown in Fig.\ref{PhaseDiagram} according to
which, at an electron temperature of $\sim$ 200mK, the cluster is radially
oriented (formation of shells) but not orientationally oriented.

As a rule, the experimental addition energies are found smaller than the
calculated ones, especially for low occupation number. This observation
probably means that the Wigner island is in an excited configuration. Its
energy is higher than that of the ground state; a lesser decrease of the
barrier is required to extract one electron. On the contrary, higher values
than the calculated ones remain unexplained (e.g., $N=5$ in
Fig.\ref{AdditionSpec2}).  The extraneous source of noise energy that seems to
be present in the experiment possibly comes from the tight coupling with the
SET. We have evidence that the SET backaction affects the temperature of the
electrons in the island in a way that depends on the SET bias current, either
heating or cooling with respect to the bath temperature. This effect is under
study.

It is known from the moving pictures of vortex lines configuration in a
superfluid rotating bucket by \citet{Williams:74} that the vortices tend to
jump around randomly. In order to damp the vortex motions so that they could
be photographed, the authors of Ref.[\onlinecite{Williams:74}] added $^3$He
impurities to the $^4$He superfluid to bring in some dissipation. In our
Wigner islands, no such dampener is introduced. Due to the extremely high
mobility of electrons over helium, it is quite likely that the electronic
configuration is also extremely unsteady.


\mbox{ }

\section{Conclusion}
 
We have studied the confinement of a small number of electrons in a trap over
a liquid helium film. Stable configurations down to a single electron can be
obtained reproducibly and their energy recorded with an SET readout. The
experimental addition spectra compare well with those obtained in Monte Carlo
simulations in a trapping potential directly derived from the cell geometry.
Charges trapped in the dielectric parts of the sample modify only weakly the
potential profile. The good agreement between actual experiments and MC
simulations 1) confirms the validity of the model (and assumptions) used in
the simulation, 2) shows that no uncontrolled, or unforeseen, feature
plays a significant role in the physical system, 3) opens the way, once the
problem of repeatability of addition spectra after corona discharges is
solved, to more detailed studies of these Wigner islands, and, in particular,
of orientational ordering.

\begin{acknowledgments}
  This work has been supported in part by Fonds National de la Science, 
  grant ACI 2002-2140. It was started as a collaboration with the Royal Holloway
  College in the framework of the European Network HPRN-CT-2000 00157.
\end{acknowledgments}

\bibliography{Wigner-islands}

\end{document}